# The human pipeline: distributed data reduction for ALMA

Scott L. Schnee[*a], Crystal Brogan[a], Daniel Espada[b,c,d], Elizabeth Humphries[e], Shinya Komugi[c], Dirk Petry[e], Baltasar Vila-Vilaro[b], and Eric Villard[b]

[a]National Radio Astronomy Observatory, 520 Edgemont Rd, Charlottesville, VA, U.S.A.;
[b]Joint ALMA Observatory, Alonso de Cordova 3107, Vitacura, 763-0355, Santiago, Chile;
[c]National Astronomical Observatory of Japan (NAOJ), 2-21-1 Osawa, Mitaka, 181-8588, Tokyo, Japan; [d]The Graduate University for Advanced Studies (SOKENDAI), 2-21-1 Osawa, Mitaka, 181-0015, Tokyo, Japan; [e]European Southern Observatory, Karl-Schwartzschild-Str. 2, 85748 Garching, Germany

## ABSTRACT

Users of the Atacama Large Millimeter/submillimeter Array (ALMA) are provided with calibration and imaging products in addition to raw data. In Cycle 0 and Cycle 1, these products are produced by a team of data reduction experts spread across Chile, East Asia, Europe, and North America. This article discusses the lines of communication between the data reducers and ALMA users that enable this model of distributed data reduction. This article also discusses the calibration and imaging scripts that have been provided to ALMA users in Cycles 0 and 1, and what will be different in future Cycles.

## 1. INTRODUCTION

The Atacama Large Millimeter/submillimeter Array (ALMA) is an instrument capable of making interferometric and total power observations of the sky from wavelengths in the range of 3 mm to 400 μm. Although its capabilities are still being expanded, ALMA is already the world's most powerful (sub)millimeter telescope. A more complete description of ALMA can be found in Hills and Beasley[1] and Wootten and Thompson[2].

Observations taken as a result of successful observing proposals undergo a series of quality assurance steps to ensure that the data delivered to principal investigators (PIs) and put into the ALMA archive meet the stated goals of projects. As part of the quality assurance process, data are calibrated and imaged, and the image properties (primarily resolution and sensitivity) are compared with those stated in proposals to ensure that the scientific goals of each project can be met. The products provided to PIs consist of the raw data, calibration and imaging scripts, calibration tables, diagnostic plots, log files, and reference images.

In Full Science operations, an automated pipeline will run the calibration and imaging steps required for quality assurance. In Early Science, before the automated pipeline is fully operational, a team of data reduction experts carries out the data processing. A description of ALMA during Early Science is described in Lundgren[3], Espada[4], and Nyman[5]. The communication between ALMA staff and ALMA users, as well as the user experience of this 'human pipeline', is described below. Other aspects of ALMA Early Science data reduction are covered in Petry[6].

---

[*] sschnee@nrao.edu, +1 434-296-0391

## 2. COMMUNICATION AND WORKLOAD DISTRIBUTION

### 2.1 Data Acquisition

The observations to be taken are defined using Scheduling Blocks (SBs), which are generated by staff at the Joint ALMA Observatory (JAO) and ALMA Regional Centers (ARCs) using the ALMA Observing Tool (OT). The ALMA OT also serves the role of proposal preparation, as described in Bridger et al.[7]. For more information on the ARCs and JAO, see Andreani et al.[8] and Nyman et al.[9]. In brief, the JAO is composed of ALMA staff based in Chile, and the ARCs are composed of ALMA staff based in North America, Europe, and East Asia. Astronomers from the JAO and ARCs carry out the observations, chosen based on the LST, weather, array configuration, proposal ranking, and executive balance between regions. This dynamic scheduling helps ensure that observations are well matched to observing conditions and that the resultant data are likely to meet the project's goals. Nevertheless, additional checks are required to determine whether or not data are of suitable quality.

### 2.2 Data Reduction Assignments

The basic unit of an ALMA observation is the Execution Block (EB), a self-contained set of observations of the science target(s) and calibrators, usually consisting of roughly 60-90 minutes of data. Based on a nominal number of antennas in the array and assumed weather conditions, the number of EBs required to meet the project goals are approximately known. This estimate can be adjusted based on the actual circumstances of the observations. The set of EBs associated with one array configuration and one spectral setup is called an Observing Unit Set (OUS). Once a sufficient number of EBs that have passed initial quality assurance checks have been collected such that the sensitivity requirement for an OUS is likely to be met, the EBs are ready to be calibrated and jointly imaged to carry out the final quality assurance checks.

In Early Science, ALMA data are calibrated and imaged by staff at the JAO and the ARCs. The North American ALMA Science Center (NAASC) has a team of about 10 staff in Charlottesville, Socorro, and Victoria that regularly carries out data reduction. In addition, most NAASC members reduce ALMA data at least once per proposal cycle to assist with quality assurance and keep current with the latest developments with ALMA. Each of the ARCs and the JAO has a Data Reduction Manager (DRM) who has the responsibility of assigning and overseeing local data reductions. The DRMs have weekly telecons to discuss new issues and common concerns, and develop a set of best practices to efficiently and accurately assess data quality.

### 2.3 Communication Between ARCs and JAO

When an OUS has the requisite number of executions that have passed initial quality assurance checks, it is marked as 'Fully Observed' and is listed on a restricted-access web page that is regularly monitored by the DRMs at the JAO and ARCs. The DRMs then assign fully observed OUSs to members of their teams, using the ALMA Helpdesk[†] and/or ALMA's issue tracking software to specify the name of the project, SB, OUS, and EBs and other details of the assignment. The DRMs also change the status of the OUS from 'Fully Observed' to 'Assigned' to avoid having the OUS undergo quality assurance at multiple sites.

Each ALMA project is associated with two tickets in JIRA[‡], our issue tracking system, one for designing the SBs and another for tracking data reduction. The data reducer for an OUS can find the science objectives for a project on the ticket for the SBs, which includes information such as the desired sensitivity, spectral resolution, and angular resolution. The data reducer reports on progress and problems with the quality assurance process on the ticket for tracking data reduction. The DRMs watch these tickets and provide guidance to the data reducers as necessary.

---

[†] https://help.almascience.org
[‡] https://www.atlassian.com/software/jira

If an OUS fails to meet the sensitivity or resolution requirements stated by the PI in the ALMA proposal, then the DRM marks the OUS as needing additional EBs and it is put back into the observing queue. If an OUS passes both the sensitivity and resolution requirements, then the DRM marks it as having passed quality assurance and approves it for delivery to the PI.

## 2.3 Communication Between ARCs and PIs

When data have passed the quality assurance checks, the PI is contacted through the ALMA Helpdesk with instructions on how to download the EBs and the reduction package from the ALMA Archive[10]. The PI is also invited to obtain additional support from the ARCs, either through the Helpdesk for simple questions or through face-to-face visits to the ARCs for more complete support.

# 3. CALIBRATION SCRIPTS

ALMA data reduction is carried out using the Common Astronomy Software Applications (CASA) package, as described in Petry[11]. Users interface with CASA via the Python programming language, making it straightforward to design and execute data reduction scripts. As observations are taken, metadata are also written to assign an 'intent' to each source observed. These metadata are then used to generate a template reduction script, using the knowledge of which targets are to be used as calibrators and which targets are to be treated as objects of scientific interest to the PI. By using the same template for reduction at the JAO and all ARCs, the project provides a standardized product to ALMA users. This also simplifies the identification of problems in the data acquisition and analysis across a distributed workforce.

## 3.1 Contents of a calibration script

A typical calibration script can be thought of as consisting of two portions, *a priori* calibration based on knowledge of the instrument and atmosphere at the time of observations and *a posteriori* calibration based on interpreting observations of calibrators. During *a priori* calibration, the reduction script 1) flags data that are known to be bad (e.g., due to an antenna being off source or a receiver becoming unlocked), 2) corrects variations of phase with respect to time (on the time scale of a few seconds to a few minutes) using measurements from water vapor radiometers[12] (present only in observations with the 12m antennas), 3) applies corrections to the positions of antennas gathered from baseline observations taken after the time of observation, and 4) applies corrections of amplitude with respect to frequency and time based on system temperature measurements.

The template calibration script also carries out basic *a posteriori* calibration and flagging. Corrections for amplitude and phase with respect to frequency are calculated using the bandpass calibrator, which is typically a regularly observed bright quasar that is observed in the same spectral setup as observations of the science target. The bandpass calibrator is selected to be a point source and during calibration is assumed to have zero spectral index. Antenna-based solutions needed to make the observations of the bandpass calibrator match the model are applied to all sources in the EB. Corrections for amplitude and phase of the science target with respect to time are calculated using the gain calibrator, which is typically the closest quasar in the sky to the science target that can be observed with sufficient signal-to-noise in about a minute of observations. The gain calibrator is selected to be a point source and antenna-based solutions needed to make the observations of the gain calibrator match the model are applied to the science target, interpolating in time. Corrections for the absolute flux scale are calculated using the flux calibrator, which is usually either a solar system object (assumed to be a uniform brightness disk) or a frequently-monitored quasar (selected to be a point source).  The absolute fluxes of all sources are bootstrapped from observations of the flux calibrator to the observations of the bandpass and phase calibrators.

The template script often has to be modified to properly calibrate an EB.  The most common modifications are additional flagging commands, for instance to remove time ranges and antennas with much larger amplitude or phase scatter than the mean.  Other common modifications include changing the number of channels to average together in bandpass solutions or changing the number of time steps to average together in gain solutions to increase the signal-to-noise for the calibration.  Less frequently, the script is modified to transfer solutions from wide spectral windows to narrow windows or change the intent of a calibrator (for instance, using a bandpass calibrator with a recently measured flux as the absolute flux calibrator due to problems with the flux calibrator).

### 3.2 Outputs of a calibration script
Along with deriving amplitude and gain solutions as functions of time and frequency, the script for calibration also makes a set of diagnostic plots of these solutions.  In addition, the data reducer generates plots of the calibrated uv-data to assist with the quality assurance process.  Data reducers review these plots to check that the calibration tables vary smoothly from channel to channel and from time step to time step.  Outliers from the general smoothness of calibration solutions indicate either data that should be flagged or solution intervals that need to be widened (in time or frequency).  Data reducers also check that the calibrated uv-data show, for the bandpass and phase calibrators, phases centered on zero degrees and amplitudes relatively flat across time and frequency.  Any antennas showing much larger scatter than the rest are flagged.  The process of running the calibration script, checking for outliers, and modifying the calibration script to reduce scatter, is iterated until the final calibrated measurement set is well-behaved.  This process can take as little as one day per EB, though the time can be much longer if the EB was observed in non-standard ways, has low signal-to-noise in any of the calibrators, or requires substantial flagging.  An experienced data reducer can calibrate multiple EBs in parallel, if required.  Once finished, running the calibration script takes a few hours on a high-performance computing node.

### 3.3 Communication related to the calibration script

If a data reducer has trouble with an assignment, questions can be posted to the to data reduction ticket in the ALMA issue tracking software, where it will be seen by the JAO and the ARC DRMs.  In addition, there is an email list for discussion of problems encountered by all data reducers at the JAO and ARCs, where help may be provided by any recipient.  Requests for changes to the manual script generator are reported in the ALMA issue tracking software and implemented by the JAO's DRM.

## 4. COMBINATION AND IMAGING

### 4.1 Combination

Although calibration is done individually for each EB, imaging is performed on the combination of all EBs in an OUS.  Therefore, in the case that there is more than one EB in an OUS, the first step in making an image is combining the uv-data taken over multiple observing sessions into a single measurement set.  If the EBs were observed close together in time (a separation of less than a week) and if the same calibrators are observed in all the EBs, then the data reducer may choose to normalize the absolute flux calibration between the EBs.  After the optional flux normalization, the EBs are concatenated together into one measurement set, which will then be used for imaging.

### 4.2 Imaging

The goal of imaging is to determine whether or not the observations meet the requirements set by the PI. During the ALMA Early Science period, data reducers produce reference images, meaning that all sources or the entire spectral range observed may not be imaged. The imaging requirements for an OUS are specified in the proposal, in which the PI identified the desired products. Depending on the science goals, the data reducer will produce continuum maps of the science targets and/or spectral line cubes. In the case of continuum maps, the data reducer will image line-free portions of the observed spectrum. In the case of a spectral line cube, the data reducer will image the portion of the observed spectrum corresponding to the line of interest, smoothing the output spectrum to the frequency width specified by the PI.

Imaging, as for calibration, is done in CASA, using the 'clean' routine for inversion from the uv-plane to the image plane. Deconvolution is also performed with the 'clean' routine, with regions of bright emission/absorption identified by eye. The data are weighted to match the resolution specified by the PI and maximize sensitivity. For most projects, this basic imaging is sufficient to determine the noise in the image, which is then compared with the sensitivity requirements set in the proposal. In some cases, the images provided to the PI can be substantially improved by carrying out additional calibration and imaging steps beyond the basic imaging described above. If required to assess the quality of the data, staff carrying out reduction occasionally employ tactics such as self-calibration on bright science targets or uv-plane continuum subtraction to image spectral lines in the presence of a strong continuum source. Producing the imaging script can take as little as one day per OUS, though the time can be much longer if self-calibration or continuum subtraction is required. Imaging of large mosaics or many spectral channels is much slower than imaging single fields and continuum maps. Once finished, running the imaging script typically takes minutes to hours, though in rare cases much more time can be required. In Cycle 0, it took an average of 71 days from the time that an OUS was marked as ready for reduction until the quality assurance process was completed. In Cycle 1, this time was reduced to 37 days.

## 5. QUALITY ASSURANCE

In addition to the checks made to ensure that each EB is properly calibrated, data reducers check the quality of images produced from the combination of all EBs in an OUS. Data reducers first check for imaging artifacts indicative of issues with calibration, e.g., bad data that need to be flagged. Next, the main criterion considered is the noise in the images. In a spectral cube, the noise is measured in a line-free channel. In a continuum map, the noise is measured far from any regions of emission or absorption. For maps made with the 12m array, the sensitivity requirement is judged to have been met if the noise is within 10-20% of that requested in the proposal. For maps made with the 7m array, the sensitivity requirement is met if the time on source and number of antennas matches that requested in the proposal. A secondary check is made to ensure that the synthesized beam size matches the angular resolution requested by the PI. Although the beam size can be adjusted through appropriate weighting of the data, in some cases additional observations in a significantly different array configuration may be required.

The data reducer makes the primary checks on the quality of the calibration and imaging. The DRMs then review all reductions and make the final decision on whether an OUS passes this level of quality assurance. If so, the OUS is marked as having passed this stage in the quality assurance process and the PI is contacted with instructions for delivery. The PI will have proprietary access to these data for one year, except in the case of time awarded through director's discretionary time, in which case the proprietary time is six months. The proprietary period is separate for each OUS. If the sensitivity goals and resolution are not met, or if there are imaging artifacts that can not be fixed, then the DRM will mark the OUS as having failed quality assurance and additional observations will be requested. The PI is not, by default, contacted in the case that the quality assurance process results in additional observations being triggered. However, the PI can track the progress of projects through the ALMA Project Tracker[13] or through queries via the ALMA Helpdesk.

If, after delivery, the PI detects problems with the data calibration and imaging, this is reported by the PI to the supporting ARC via the ALMA Helpdesk. The DRM at the ARC receiving the Helpdesk ticket has the responsibility of investigating the PI's concerns. If confirmed, the DRM opens a new ticket in the ALMA issue tracking software to alert the DRMs at the JAO and other ARCs. The DRMs at all sites work together to determine the scope of the issue (affecting a small number of projects or many projects) and a solution to the issue (fixes or re-observation). Deliveries and downloads of all OUSs potentially affected by the issue are suspended during this period of investigation. Once a solution is implemented, the PI of all affected projects are contacted by the ARCs via the ALMA Helpdesk with information and, as soon as possible, corrected data. The proprietary periods of all affected projects will be adjusted to reflect the time spent investigating and solving the problem.

Because the quality assurance process does not require that every science target be imaged across the entire range of observed frequencies, the PI may want to carry out additional imaging. The imaging script produced by the data reducer provides a useful starting point for the PI, and there are many CASA Guides on the web[§] that provide an easy reference. Should additional help be required, ALMA users are encouraged to use the ALMA Helpdesk to get relatively simple questions answered or to schedule face-to-face visits for help with more complex issues.

## 6. FUTURE

As ALMA develops new capabilities, the data reduction process will continue to evolve. Total power (single dish) observations were offered in Cycle 1 and will be carried over to Cycle 2, requiring new calibration and imaging routines to be implemented. In Cycle 2, projects will include spectral scans and polarization observations. These observing modes will need to be supported by either manual or pipeline calibration and imaging procedures, or both. The most fundamental upcoming change will be the transition to pipeline-assisted reductions. In Cycle 2, the automatic pipeline is expected to be able to calibrate the majority of the observations, leaving manual reducers to handle all imaging and some of the trickiest calibrations. In later Cycles, the automatic pipeline will be able to handle a larger fraction of calibration and will be able to produce reference images of sufficient quality to be used for quality assurance. An eventual goal of the pipeline is to produce science-ready maps and spectral cubes. Nevertheless, it is expected that by-hand reductions will be required for the foreseeable future.

## 7. SUMMARY

ALMA observations are taken by staff in Chile, using SBs written by staff at the Joint ALMA Observatory and ALMA Regional Centers. When an SB has been executed enough times to meet the sensitivity goals of the PI, these data are marked as ready for reduction. In ALMA Early Science Cycles 0 and 1, all data for PIs were calibrated and imaged by a team of data reduction experts at the JAO and ARCs. Using a common template and making changes as needed, data reducers check that the observations can be calibrated successfully and that the resultant images meet the resolution and noise goals described in the observing proposal. When observations pass this level of quality assurance, the data are archived and the PIs are contacted by ARCs and given instructions on how to download the data. When observations fail the quality assurance process, the JAO puts the SB back in the queue to trigger further observations. PIs use the ALMA Helpdesk to communicate with the ARC. Intra-ARC communication between data reducers and DRMs takes place using the ALMA Helpdesk and the issue tracking software, as well as less formal methods. Communication between the ARCs and the JAO takes place using wiki and web pages, email lists, issue tracking software, and weekly telecons.

## 8. ACKNOWLEDGEMENTS

---

[§] http://casaguides.nrao.edu/